\documentclass[12pt]{iopart}
\usepackage{graphicx}
\eqnobysec
\newcommand{\beq}{\begin{equation}}
\newcommand{\beqa}{\begin{eqnarray}}
\newcommand{\eeq}{\end{equation}}
\newcommand{\eeqa}{\end{eqnarray}}
\newcommand{\bigmean}[1]{\left\langle#1\right\rangle}
\newcommand{\cusp}{{\rm cusp}}
\renewcommand{\d}{{\rm d}}
\newcommand{\ds}{\displaystyle}
\renewcommand{\e}{{\rm e}}
\newcommand{\eps}{\varepsilon}
\newcommand{\euler}{\gamma}
\newcommand{\f}{\varphi}
\newcommand{\g}{\psi}
\renewcommand{\hat}{\widehat}
\renewcommand{\i}{{\rm i}}
\newcommand{\kes}{{({\rm Kes})}}
\renewcommand{\l}{{\lambda}}
\newcommand{\mean}[1]{\langle#1\rangle}
\newcommand{\n}{{\nu}}
\newcommand{\p}{{(p)}}
\newcommand{\prob}{\mathop{\rm Prob}\nolimits}
\newcommand{\q}{q}
\newcommand{\stir}[2]{\left[#1\atop#2\right]}
\renewcommand{\tilde}{\widetilde}
\newcommand{\var}{\mathop{\rm var}\nolimits}
\newcommand{\A}{{\cal A}}
\newcommand{\B}{{\cal B}}
\newcommand{\D}{{\Delta}}
\newcommand{\Int}{\mathop{\rm Int}\nolimits}
\newcommand{\LL}{{\cal L}}
\newcommand{\N}{{\cal N}}
\renewcommand{\Re}{\mathop{\rm Re}\nolimits}
\renewcommand{\SS}{{\cal S}}

\begin{document}

\title{A record-driven growth process}

\author{C Godr\`eche and J M Luck}

\address{
Institut de Physique Th\'eorique, IPhT, CEA Saclay, and URA 2306, CNRS,
91191 Gif-sur-Yvette cedex, France}

\begin{abstract}
We introduce a novel stochastic growth process, the {\it Record-driven growth
process}, which originates from the analysis of a class of growing networks in
a universal limiting regime.
Nodes are added one by one to a network, each node possessing a quality.
The new incoming node connects to the preexisting node with best quality,
that is, with record value for the quality.
The emergent structure is that of a growing
network, where groups are formed around record nodes (nodes endowed with the
best intrinsic qualities).
Special emphasis is put on the statistics of leaders (nodes whose degrees are
the largest).
The asymptotic
probability for a node to be a leader is equal to the Golomb-Dickman constant
$\omega=0.624\,329\dots$,
which arises in problems of combinatorical nature.
This outcome solves the problem of the determination of the record breaking
rate for the sequence of correlated inter-record intervals.
The process exhibits temporal self-similarity in the late-time regime.
Connections with the statistics of the cycles of random permutations, the
statistical properties of randomly broken intervals, and the Kesten variable
are given.
\end{abstract}


\eads{\mailto{claude.godreche@cea.fr},\mailto{jean-marc.luck@cea.fr}}

\maketitle

\section{Introduction}
\label{intro}

In the present work we introduce and study a novel stochastic growth process,
the {\it Record-driven} (RD) {\it growth process},
which originates from the analysis
of a class of growing networks in a universal limiting regime.
As we shall see, this process is also related to three other fields, namely the
statistics of the cycles of random permutations, the statistical properties of
randomly broken intervals, and the Kesten variable.

The RD growth process originates
from a class of growing networks with a preferential attachment rule,
amongst which the most well known representative is the model of Barab\'asi and
Albert (BA)~\cite{ba}.
The latter provides a natural explanation for the main features
observed in real networks~\cite{abrmp,doro,blmch,cup}, and chiefly their
scale-freeness,
testified by the power-law fall-off of their degree distribution.
Relevant for our purpose are networks where the attachment rule involves both
the degree of the nodes and their intrinsic quality or fitness, an extension of
the original BA model which is due to
Bianconi and Barab\'asi (BB)~\cite{bb}.
The BB model has the remarkable feature that it exhibits a continuous
condensation transition, analogous to the Bose-Einstein condensation.

A general definition of this class of networks is as follows.
The network being initially empty, at each integer time step, $n=1,2,\dots$, a
new node, labeled by its birth date~$n$, is added.
Node $n$ is endowed once and for all with an intrinsic quality $\eta_n$,
modeled by a quenched random variable drawn from some given distribution.
This node (except for the first one) connects by one link
to any of the earlier nodes ($i=1,\dots,n-1$) with probability
\beq\label{pin}
p_{i,n}=\frac{\eta_i\, h(k_i)}{Z_{n-1}},
\eeq
where $h(k_i)$ is a function of the degree $k_i$ of node~$i$,
i.e., the number of nodes already connected to node $i$ at time $n-1$.
The denominator
\beq
Z_{n-1}=\sum_{i=1}^{n-1}\eta_i\,h(k_i)
\eeq
is a normalization factor.
The probability $p_{i,n}$ is thus proportional to
an intrinsic factor, the quality $\eta_i$ of node $i$, and to
a dynamical one, represented by $h(k_i)$.
Preferential attachment to nodes whose degree is already larger
is realized whenever the function $h(k_i)$
is an increasing function of the degree $k_i$.
Finally, temperature $T$ is introduced into the model by
considering the qualities as activated variables, i.e., by setting
\beq
\eta_n=\exp(-\eps_n/T),
\eeq
and assuming that the activation energies $\eps_n$
are drawn from a given temperature-independent continuous distribution.

The BA and BB models in their original forms
both correspond to the linear function $h(k)=k$.
The BA model corresponds to the limiting case of infinite temperature.

After time step $n$ is completed
the network consists of $n$ nodes connected by $n-1$ links.
It therefore has the topology of a tree.
In the rest of the paper we use a slightly modified definition of the degree,
keeping abusively the same notation, $k_i$.
We define the degree of a given node as the number of {\it incoming} links on
this node,
taking aside its unique outgoing link, except for the first node, which has
no outgoing link.
This new definition corresponds to shifting by one unit the former quantity,
$k_i\to k_i-1$, except for the first node.
The degrees thus defined sum up to the number of links, i.e.,
\beq
\sum_{i=1}^n k_i=n-1.
\eeq

As we now explain, the RD growth process is the zero-temperature limit
of the class of models described above.
In this limit, at any time $n$
the node $i$ with the lowest energy:
\beq
\eps_i=\min(\eps_1,\dots,\eps_{n-1}),
\label{besti}
\eeq
i.e., with the highest quality, has an attachment probability~$p_{i,n}$
which is overwhelmingly larger than all the other ones,
since the ratio $p_{i,n}/p_{j,n}$ grows exponentially at low temperature,
as $\exp((\eps_j-\eps_i)/T)$.
Every new node $n$ therefore connects to the earlier node $i$
with best quality at time $n$, given by~(\ref{besti}).
The successive best qualities are known as records~\cite{ren,glick,rec1,rec2}.
The corresponding nodes, that we term {\it record nodes},
are therefore the only ones whose degree grows.
This process is universal in a very strong sense.
It is independent of the function $h(k)$
and of the distribution of the node energies,
provided the latter is continuous,
so that the event $\eps_i=\eps_j$ with $i\ne j$ has zero probability.
In particular, the RD growth process is independent of whether the model has a
condensation transition or not.

Let us summarize the definition of the process and give a pictorial
interpretation of it.
Nodes (or individuals) arrive one by one to form a network of relationships.
The network being initially empty, at each integer time step, $n=1,2,\dots$, a
new node, labeled by its birth date~$n$, is added,
and endowed once and for all with an intrinsic quality $\eta_n$.
This node connects by one link to the earlier node with best quality.
This directed link can be pictorially described as ``being a disciple of''.
Thus groups are formed around record nodes (or pictorially ``masters'').
The size of a group, or the degree of a record node, is the number of incoming
links to it (i.e., the number of disciples).
Times at which a node appears with a quality that breaks the previous record
are {\it record times}.
Finally, there is another, yet simpler description of the
process relying on records only: the record times are the dates of birth of the
record nodes; the degree of the newly born record node grows linearly in time,
then stops growing when the next record node is born.

In the present work
the main emphasis will be put on the interplay between {\it records}
and {\it leaders}.
While a node is a record if its quality (an intrinsic property)
is better than those of all the earlier nodes,
a node will be said to be a leader at a given time
if its degree (a dynamical, time-dependent quantity)
is larger than those of all the other nodes.
Investigating the interplay between records and leaders is natural in the low
temperature regime of the BB model because the question is whether the best
fitted node will or will not be the leader in the course of time.
The statistics of leaders and of lead changes has been addressed previously for
the BA and related models in~\cite{kr}.
In the present case these subtle questions
can be analytically approached and given a comprehensive quantitative answer.

The main outcomes of this paper are the following.
We first determine the invariant measure associated with
the dynamical system~(\ref{rec}).
This measure gives the asymptotic distribution of the fraction of nodes
connected to a leader,
or alternatively that of the longest inter-record interval.
The knowledge of the invariant measure allows the computation of the
probability for a node to be a leader, or equivalently the probability of a
record breaking for the sequence of inter-record intervals.
We find that this probability is equal to the Golomb-Dickman constant
$\omega=0.624\,329\dots$, eq.~(\ref{omega}).
We then perform the analysis of the statistics of the difference of the labels
of two successive leaders (equivalently, of two successive records for the
sequence of inter-record intervals), as well as a thorough study of the
statistics of the lengths of time associated with the reign of a leader.
We finally explain the connections of this process with the statistics of the
cycles of random permutations, the statistical properties of randomly broken
intervals, and the Kesten variable.

The bulk of the paper begins at section~\ref{process}.
In the next section we first establish the needed background knowledge on
records,
since the latter play a fundamental role in the definition of the process under
study.

\section{Statistics of records}
\label{records}

The discrete theory of records is classical~\cite{ren,glick,rec1,rec2}.
The continuum theory, which is instrumental in our work, is less documented,
as is its relationship to a renewal process.

\subsection{Discrete theory}
\label{records:discrete}

Given a sequence of numbers, $\q_1,\q_2,\dots$,
the value $\q_i$ is a {\it record} if it is larger than all previous ones:
\beq
\q_i>\max(\q_1,\dots,\q_{i-1}).
\eeq
These numbers are for example the successive observations of a random signal,
or the successive drawings of a random variable,
modeled as a sequence of independent identically distributed (i.i.d.)
random variables.
In order to avoid ambiguities due to ties, the distribution of the latter is
taken continuous.
In the present context, the $\q_i$ stand for the qualities $\eta_i$ of the
nodes.

Referring to the label $i=1,2,\dots$ as time,
the time~$i$ of the occurrence of a record is a {\it record time}.
The definition of a record involving only inequalities between
the variables~$\q_i$, the statistics of record times is independent of
the underlying distribution of these variables.

The first value $\q_1$ is always a record.
The occurrence of a record at any subsequent time $i\ge2$
has probability $1/i$:
\beq
\prob(\q_i>\max(\q_1,\dots,\q_{i-1}))=\frac{1}{i}.
\label{1i}
\eeq
This holds independently of the occurrence of any other record,
either at earlier times $(j<i)$ or at later times $(j>i)$.
Indeed, $\q_1$ is a record with probability 1.
The probability that $\q_2$ be a
record, i.e., be larger than $\q_1$, is obviously equal to $1/2$.
The probability that $\q_3$ be a record is equal to $1/3$ because it can be
smaller, intermediate or larger than the two previous values $\q_1$ and~$\q_2$,
with equal probability, and so on.
Eq.~(\ref{1i}) can alternatively be recovered by noticing that,
amongst the $i!$ permutations of $\q_1,\dots,\q_i$,
there are $(i-1)!$ permutations where $\q_i$ is the largest.

Eq.~(\ref{1i}) means that the rate of record breaking is equal to $1/i$,
or otherwise stated that the occurrence of a record is a Bernoulli process with
success probability equal to~$1/i$.
The indicator variables $I_i$,
equal to 1 if $\q_i$ is a record and to 0 otherwise,
will be the building blocks for
the derivation of the results of this section.
It is easy to convince oneself that they are independent (for a proof,
see~\cite{ren,glick,rec1,rec2}).
We will first determine the distribution of
the number of records up to time $n$, then that of the record times.

The number $M_n$ of records up to time $n$
is a random variable taking the values $1,\dots,n$,
which can be expressed as
\beq
M_n=I_1+I_2+\cdots+I_n.
\eeq
Its average and variance read
\beq
\mean{M_n}
=\sum_{i=1}^n\mean{I_i}
=\sum_{i=1}^n\frac{1}{i}
\approx\ln n+\euler,
\label{mmean}
\eeq
\beq
\var M_n
=\sum_{i=1}^n\mean{I_i(1-I_i)}
=\sum_{i=1}^n\left(\frac{1}{i}-\frac{1}{i^2}\right)
\approx\ln n+\euler-\frac{\pi^2}{6}.
\label{mvar}
\eeq
where $\euler=0.577\,215\dots$ is Euler's constant.
More generally, the generating function of~$M_n$ reads
\beq
\mean{x^{M_n}}
=\prod_{i=1}^n\mean{x^{I_i}}
=\prod_{i=1}^n\left(1+\frac{x-1}{i}\right)
=\frac{x(x+1)\dots(x+n-1)}{n!}.
\label{gen}
\eeq
The `rising power' appearing in the right-hand side of~(\ref{gen})
is known to be the generating function of the Stirling numbers
of the first kind~\cite{knuth}:
\beq
x(x+1)\dots(x+n-1)=\frac{\Gamma(x+n)}{\Gamma(x)}
=\sum_{m=1}^n\stir{n}{m}x^m.
\label{xstir}
\eeq
The Stirling number of the first kind $\stir{n}{m}$
is the number of ways of arranging $n$ objects in~$m$ cycles,
or the number of permutations of $n$ objects having $m$ cycles,
hence
\beq
\sum_{m=1}^n\stir{n}{m}=n!,
\label{sstir}
\eeq
consistently with~(\ref{xstir}), setting $x=1$.
The distribution of~$M_n$ is therefore given by
\beq
\prob(M_n=m)=\frac{\stir{n}{m}}{n!}.
\label{prob:mn}
\eeq
In other words, the number $M_n$ of records up to time $n$
is distributed as the number of cycles in a random permutation on $n$ elements
(see section~\ref{connection1} for further developments on the connection of
records with random permutations).

The asymptotic form of the Stirling numbers
\beq
\stir{n}{m}\approx\frac{(n-1)!\,(\ln n)^{m-1}}{(m-1)!},
\label{astir}
\eeq
is readily derived from~(\ref{xstir}) by simplifying
the gamma functions for large $n$ and small~$x$
as $\Gamma(x+n)\approx(n-1)!\,n^x$ and $\Gamma(x)\approx1/x$.
As a consequence, the asymptotic form of the distribution of $M_n$ reads
\beq\label{eq:mn}
\prob(M_n=m)\approx\frac{1}{n}\frac{(\ln n)^{m-1}}{(m-1)!}.
\label{amn}
\eeq
The bulk of the distribution of $M_n$ is thus, up to a shift by one unit,
asymptotically given by a Poissonian law of parameter $\ln n$.
This holds only to leading order in $\ln n$.
The exact large-$n$ behavior
of the mean and variance of $M_n$, eqs.~(\ref{mmean}) and~(\ref{mvar}), differs
by additive constants
from the predictions of~(\ref{eq:mn}),
$\mean{M_n}_{\rm Poisson}=\ln n+1$, $(\var{M_n})_{\rm Poisson}=\ln n$.

Let us denote by $N_1,N_2\dots,N_m,\dots$ the successive record times.
Thus $N_1=1$, since~$\q_1$ is always a record.
Then, if $\q_{i_2}$ is the next record after
$\q_1$, the second record time is $N_2=i_2$, etc.
The distribution of the $m$--th record time $N_m$ follows from~(\ref{prob:mn}):
\beq
\prob(N_m=k)=\frac{\stir{k-1}{m-1}}{k!}\quad(k=m,m+1,\ldots),
\label{prob:nm}
\eeq
and $m\ge2$.
We have indeed the equivalence of events
\beq
\{N_m=k\}=\{M_{k-1}=m-1,M_k=m\}=\{M_{k-1}=m-1,I_k=1\},
\eeq
where the two events appearing inside the rightmost brackets are independent.
Using the asymptotic form~(\ref{astir}) one gets
\beq
\prob(N_m=k)\approx\frac{1}{k^2}\frac{(\ln k)^{m-2}}{(m-2)!}
\label{anm}
\eeq
for large values of $k$ and $m$.

The sequence $N_1,N_2\dots,N_m,\dots$ of record times
can be generated recursively.
We have
\beq
\prob(N_m=k\vert N_{m-1}=j)\equiv p_{k\vert j}=\frac{j}{k(k-1)},
\label{pkj}
\eeq
independently of $m$, i.e., independently of the occurrence of earlier records.
Indeed,
\beqa
p_{k\vert j}&=&\prob(I_{j+1}=\cdots=I_{k-1}=0,I_k=1)\nonumber\\
&=&\prob(I_{j+1}=0)\dots\prob(I_{k-1}=0)\prob(I_k=1)\nonumber\\
&=&\frac{j}{j+1}\,\frac{j+1}{j+2}\cdots\frac{k-2}{k-1}\,\frac{1}{k},
\label{pkjderiv}
\eeqa
because of the statistical independence of the elementary events.
The product of fractions simplifies to~(\ref{pkj}).
Consequently,
\beq
\prob(N_m\ge k\vert N_{m-1}=j)
=\sum_{l\ge k}p_{l\vert j}=\frac{j}{k-1},
\eeq
or else
\beq
\prob(N_m\ge k\vert N_{m-1})=\frac{N_{m-1}}{k-1}.
\eeq
This expression can be recast in the form of the random recursion
\beq
N_m=1+\Int\frac{N_{m-1}}{U_m},
\label{nm}
\eeq
where $\Int x$ denotes the integer part of the real number $x$,
and $U_m$ is a uniform random variable between 0 and 1,
independent of $N_{m-1}$.

We now look at the distribution of record times from a different viewpoint.
For a fixed instant of time $n\ge1$,
we consider the time $N_M$ of occurrence of the last record before $n$
($n$ being included),
that is the $M_n$--th record\footnote{In the following we drop the subscript
when it is attached
to a quantity which is itself in a subscript, whenever there is no ambiguity.}.
The time~$N_M$ is a discrete random variable uniformly distributed between 1
and~$n$:
\beq
\prob(N_M=k)=\frac{1}{n}\quad(1\le k\le n).
\label{prob:nM}
\eeq
Indeed, the joint distribution of $N_M$ and $M_n$ reads
\beqa
\prob(N_M=k,M_n=m)
&=&\prob(I_{k+1}=\dots=I_n=0)\prob(N_m=k)\nonumber\\
&=&\frac{k}{n}\,\prob(N_m=k).
\label{md}
\eeqa
for $m\le k\le n$.
Using~(\ref{prob:nm}), (\ref{sstir}) and summing~(\ref{md}) over $m$
yields the result.
Eq.~(\ref{prob:nM})
can be alternatively recovered by noticing that the time $N_M$ is entirely
characterized by the property that
the signal~$\q_i$ takes its maximal value over $i=1,\dots,n$ at time $i=N_M$.
These $n$ possible values of $N_M$ are clearly equally probable.

Similarly, consider the time $N_{M+1}$ of occurrence of
the $(M_n+1)-$st record, i.e., of the first record after a given time $n$.
It is easily found, by the same reasoning leading to (\ref{pkj}), that
\beq
\prob(N_{M+1}=k)=\frac{n}{k(k-1)}\quad(k\ge n+1).
\eeq
Therefore
\beq
\prob\left(\frac{n}{N_{M+1}}<\frac{n}{k}\right)=\frac{n}{k},
\label{prob:nMp}
\eeq
from which one can infer that, when $n\to\infty$, $n/N_{M+1}$ is uniformly
distributed between~0 and 1.
This will be proved below, by another method.

\subsection{Continuum theory}
\label{records:continuum}

At large times, the discrete theory of records
has an asymptotically exact description in terms of continuous variables.
This description stems from the recursion, inherited from~(\ref{nm}),
\beq
N_m=\frac{N_{m-1}}{U_m},
\label{rec:cont}
\eeq
where the $N_m$ are now considered as real numbers.
The product structure of this recursion suggests to introduce
a logarithmic scale of time.
In this time scale the process of record times
is transformed into a very simple renewal process~\cite{feller,cox,cox-miller},
as shown below.
This remark allows an easy access to the determination of the asymptotic
distributions of the observables needed in the sequel ($M_n$, $N_m$, $N_M$).

We set\footnote{The shift of the index $m$ by one unit is due to the fact
that there is a record at $N_1=1$,
while the usual convention $t_0=0$ holds in renewal theory.}
\beq
t=\ln n,\quad t_m=\ln N_{m+1},\quad\tau_m=-\ln U_{m+1}.
\eeq
The $U_m$ being uniformly distributed between 0 and 1, the increments $\tau_m$
are i.i.d.~random variables
with density $\rho(\tau)=\e^{-\tau}$.
The recursion~(\ref{rec:cont}) becomes
\beq
t_m=t_{m-1}+\tau_m,
\label{rect}
\eeq
which defines the time of occurrence of the $m$--th renewal,
\beq
t_m=\tau_1+\cdots+\tau_m.
\eeq
Its probability density $f_{t_m}$ is therefore equal to
the $m$--th convolution product of $\rho$, corresponding to the gamma
distribution:
\beq
f_{t_m}(T)=\e^{-T}\frac{T^{m-1}}{(m-1)!},
\eeq
from which we formally deduce the density of $N_{m+1}=\exp(t_m)$
\beq
f_{N_{m+1}}(N)=\frac{1}{N^2}\frac{(\ln N)^{m-1}}{(m-1)!},
\label{fnm}
\eeq
in agreement with the asymptotic form~(\ref{anm})
of the law of $N_{m}$.
This is the law of the inverse of the product of $m$ uniform random
variables between 0 and 1.

The number of records up to time $n$, $M_n$, translates into
the number $\N_t$ of renewals up to time $t$, up to a shift by one unit
between the two quantities,
\beq
M_n=\N_t+1.
\eeq
In the present case of exponentially distributed increments,
the distribution of $\N_t$ is Poissonian:
\beq
\prob(\N_t=m)=\e^{-t}\frac{t^m}{m!},
\label{npoiss}
\eeq
in agreement with the asymptotic form~(\ref{amn}) of the law of $M_n$.

Consider finally the ratios
\beqa
&&X_n=\frac{N_M}{n}=\exp(-B_t),\quad
Y_n=\frac{N_{M+1}}{n}=\exp(E_t),\nonumber\\
&&Z_n=\frac{N_{M+1}}{N_M}=\frac{Y_n}{X_n}=\exp(B_t+E_t),
\label{xyzdef}
\eeqa
where the backward and forward recurrence times are respectively defined as
\beq
B_t=t-t_\N,\quad E_t=t_{\N+1}-t.
\eeq
For a renewal process with exponentially distributed increments,
the latter quantities are statistically independent
of each other and have the same exponential distribution
as the increments~\cite{gl}, i.e.,
\beq
f_{B_t}(B)=\e^{-B},\quad f_{E_t}(E)=\e^{-E}.
\label{fbe}
\eeq
We can thus conclude that $X_n$ is uniform between 0 and 1,
consistently with~(\ref{prob:nM}),
that~$Y_n$ is equal to the inverse of such a uniform random variable,
consistently with what was suggested by~(\ref{prob:nMp}),
and that $Z_n$ is therefore equal to the inverse of a product of two such
uniform random variables, namely
\beqa
&&f_X(x)=1\quad(0<x<1),\qquad
f_Y(y)=\frac{1}{y^2}\quad(y>1),\nonumber\\
&&f_Z(z)=\frac{\ln z}{z^2}\quad(z>1).
\label{xyzlaw}
\eeqa

\section{The Record-driven growth process}
\label{process}

The definition of the process given in the Introduction can now be made more
quantitative, first at the level of the discrete formalism, then in the
continuum limit.

\subsection{Discrete description}
\label{process:discrete}

Let $k_i(n)$ denote the degree of node~$i$ (defined as the number of incoming
links on this node) at the later time $n\ge i$.
We have clearly $k_i(i)=0$.
Then:

\noindent $\bullet$
If $\eta_i$ is not a record, i.e., if the time $i$ is not a record time,
the degree of node $i$ does not grow any further:
\beq
k_i(n)=0\;\;\hbox{for all}\;\;n\ge i.
\label{k1}
\eeq

\begin{figure}[!ht]
\begin{center}
\includegraphics[angle=90,width=.5\textwidth]{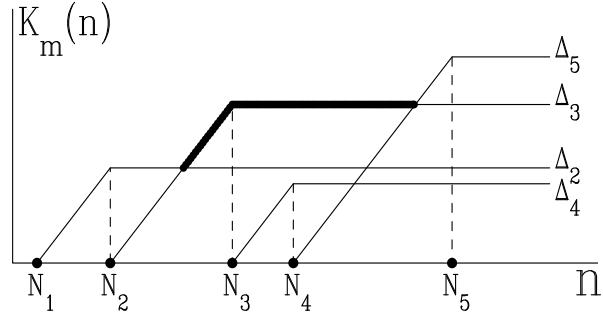}
\caption{
Schematic plot of the time dependence of the degrees $K_m(n)$
of the first few record nodes ($m=1,\dots,4$)
in the RD growth process.
The thick line shows the part of its history where the second record node
($m=2$) is the leader.}
\label{mod}
\end{center}
\end{figure}

\noindent $\bullet$
If $\eta_i$ is a record,
i.e., if the time $i=N_m$ belongs to the sequence of record times,
the degree of node $i$ grows linearly with time
until the subsequent record is born at time~$N_{m+1}$.
It then stays constant.
Setting
\beq
i=N_m,\quad k_i(n)=K_m(n),
\label{k1k}
\eeq
one has therefore for the degree of the $m$--th record node
\beq
K_m(n)=\left\{\matrix{
n-N_m\hfill&(N_m\le n\le N_{m+1}),\cr
N_{m+1}-N_m\quad&(n\ge N_{m+1}).\hfill
}\right.
\eeq
The limit value of $K_m(n)$, that is $K_m(N_{m+1})$,
is simply given by the inter-record interval, which we denote by
\beq
\D_{m+1}=N_{m+1}-N_m.
\label{dfond}
\eeq
Figure~\ref{mod} illustrates these definitions.

At any given instant of time $n$,
there is a node whose degree is larger than that
of all existing nodes at this time.
We term this node {\it the leader at time $n$}.
We will also say that a given node {\it is a leader}
if it leads for some period of time during its history.
For instance, on Figure 1, the record nodes 1, 2 and 4 are leaders,
but 3 is not.
If $m$ denotes the number of record nodes before time $n$,
i.e., $N_m\le n<N_{m+1}$, the degree of the leader at time $n$ reads
\beqa
L(n)&=\max(K_1(n),K_2(n),\dots,K_m(n))\nonumber\\
&=\max(\D_2,\dots,\D_m,n-N_m)\nonumber\\
&=\max(L_m,n-N_m),
\label{fund}
\eeqa
where
\beq
L_m=\max(\D_2,\dots,\D_m)
\label{ldef}
\eeq
is the degree of the leader, $L(N_m)$, at the record time $N_m$ ($m\ge2$).

For the time being we focus our attention
onto the degree of the leader at record times.
We will return to the case of a generic time in section~\ref{secgeneric}.
Eq.~(\ref{fund}) yields the fundamental recursion relation
\beq
L_{m+1}=\max(L_m,\D_{m+1}).
\label{fond}
\eeq
The meaning of this recursion is as follows.

\noindent $\bullet$
If $\D_{m+1}>L_m$, the $m$--th record node is the leader at time $N_{m+1}$.
One has therefore $L_{m+1}=\D_{m+1}=K_m(N_{m+1})$.

\noindent $\bullet$
If $\D_{m+1}\le L_m$,
the $m$--th record node is not the leader at time $N_{m+1}$.
The degree of the leader is left unchanged, i.e., $L_{m+1}=L_m$.
It turns out that the $m$--th record node will never be a leader
(see~(\ref{leadineqs})).
This node is said for short to be a {\it subleader}.

The condition $\D_{m+1}>L_m$ is equivalent to saying that $\D_{m+1}$ is
larger than all the previous $\D_k$ for $k=2,\dots,m$,
i.e., {\it is a record for the sequence of
inter-record times} $\D_2,\D_3,\dots$
Leaders therefore correspond to records for this sequence.
On the example of Figure~\ref{mod}
we have $L_2=\Delta_2<L_3=L_4=\Delta_3<L_5=\Delta_5$, but $\Delta_3>\Delta_4$.
Thus the record nodes 1, 2, and 4 are leaders, and $\Delta_2$, $\Delta_3$ and
$\Delta_5$ are records.

Let
\beq\label{omegm}
\omega_m=\prob(\D_{m+1}>L_m)
\eeq
denote the probability for the $m$--th record node be a leader, or
as mentioned above, the probability of a record breaking at step $m+1$
for the sequence of inter-record intervals $\D_2,\D_3,\dots$
In section~\ref{onetime} we will determine the limit $\omega$ of $\omega_m$
when $m\to\infty$, that is, the
probability for a record node to be a leader,
in the limit of long times.
At this stage we can give the expression of this quantity in
the first two steps of the process.
One has clearly $\omega_1=1$.
Furthermore it is easy to see that the joint distribution
of the record times $N_2$ and $N_3$ reads
\beq
\prob(N_2=j,N_3=k)=p_{j\vert1}\,p_{k\vert j}
=\frac{1}{(j-1)(k-1)k}\quad(k>j>1).
\eeq
For $N_2=j$ and $N_3=k$, we have $L_1=0$, $L_2=\D_2=j-1$, and $\D_3=k-j$.
The first record node, born at time $N_1=1$, is always a leader.
The second record node, born at time $N_2=j$,
is a leader if $\D_3>L_2$, i.e., $k\ge2j$.
This occurs with probability
\beqa
\omega_2
&=&\sum_{j\ge2}\sum_{k\ge2j}\frac{1}{(j-1)(k-1)k}
=\sum_{j\ge2}\frac{1}{(j-1)(2j-1)}
\nonumber\\
&=&2(1-\ln2)=0.613\,705\dots
\label{w2}
\eeqa
The expressions of $\omega_m$ become increasingly complex as $m$ becomes larger
and larger.
Let us mention the following result without proof:
\beqa
\omega_3&=&{\rm Li}_2\left(\frac{1}{4}\right)-\frac{7\pi^2}{36}
-\frac{\pi\sqrt{3}}{12}+3(\ln 2)^2-\frac{3\ln3}{4}-2\ln2+\frac{7}{2}
\nonumber\\
&=&0.626\,218\dots,
\label{w3}
\eeqa
where the first term involves the dilogarithm function.

\begin{figure}[!ht]
\begin{center}
\includegraphics[angle=90,width=.5\textwidth]{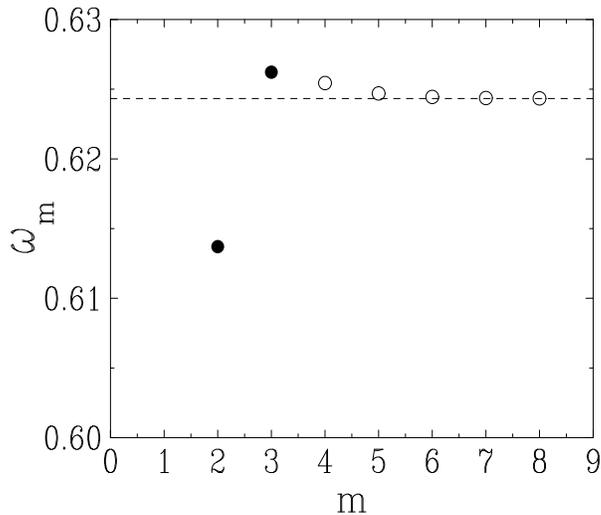}
\caption{
Plot of the probability $\omega_m$ that the $m$--th record node is a leader,
against~$m$.
Full symbols: exact values~(\ref{w2}) and~(\ref{w3}).
Empty symbols: data for higher $m$, obtained by a numerical simulation
based on the recursions~(\ref{nm}), (\ref{dfond}) and~(\ref{fond}).
Dashed line: asymptotic limit given by the Golomb-Dickman constant $\omega$,
eq.~(\ref{omega}).}
\label{wn}
\end{center}
\end{figure}

We performed an accurate numerical evaluation of the $\omega_m$
by a simulation of the process based on a generation of records using
eqs.~(\ref{nm}), (\ref{dfond}) and~(\ref{fond}).
Figure~\ref{wn} demonstrates the very fast convergence
of the $\omega_m$ to the limit $\omega$,
known as the Golomb-Dickman constant (see~(\ref{omega})).
In contrast with the case of i.i.d.~random variables, the record breaking rate
of the sequence of $\Delta_m$ goes to a non-trivial constant.

Denoting by $\nu=1,2,\dots$ the difference of the labels of any two successive
leaders, say $m$ and $m+\nu$,
the ``reign'' of the leader born at $N_m$, i.e., the period of time during which
it stays a leader, begins at time
\beq
a_m=N_m+L_m
\label{cdef}
\eeq
and ends at time
\beq
b_m=a_{m+\nu}=N_{m+\nu}+L_{m+\nu}.
\label{adef}
\eeq
The reign therefore has a duration
\beq
l_m=b_m-a_m=a_{m+\nu}-a_m.
\label{tdef}
\eeq
The statistics of these times is addressed in section 5.

Finally, the inequalities
\beq
a_m<N_{m+1}<b_m
\label{leadineqs}
\eeq
prove that the $m$--th record node
is a leader (for some period of time) if and only if it leads at time $N_{m+1}$,
i.e., if and only if $\D_{m+1}>L_m$.

\subsection{Continuum theory}
\label{process:continuum}

As stated previously,
the statistics of records is faithfully described in the regime of long times
by a continuum approach.
The late stages of the RD growth process
have a similar asymptotically exact continuum description.

In the continuum theory the key quantity is the ratio
\beq
R_m=\frac{L_m}{N_m},
\label{rdef}
\eeq
which is the fraction of nodes which are connected to the leader at that time.
Indeed the numerator $L_m$ is the degree of the leader at time $N_m$,
while $N_m$ is equal to the number of nodes in the system.
According to~(\ref{ldef}), $R_m$ is also the scaled maximal inter-record interval.
The values taken by $R_m$ are clearly between 0 and 1.
Recalling~(\ref{rec:cont}), the recursion~(\ref{fond}) becomes
\beq
R_{m+1}=\max(U_{m+1}R_m,\,1-U_{m+1}),
\label{rec}
\eeq
where $U_{m+1}$ is uniformly distributed between 0 and 1 and independent of
$R_m$.
The branches of~(\ref{rec}) correspond respectively to the events
\beqa
\LL_m=\{m\hbox{ leader}\}\hfill&=&\left\{U_{m+1}<\frac{1}{1+R_m}\right\}
\Rightarrow R_{m+1}=1-U_{m+1},\nonumber\\
\SS_m=\{m\hbox{ subleader}\}&=&\left\{U_{m+1}>\frac{1}{1+R_m}\right\}
\Rightarrow R_{m+1}=U_{m+1}R_m.\nonumber\\
\label{leader}
\eeqa

The stochastic dynamical system defined by (\ref{rec})
plays a central role in the following.
It is reminiscent of the recursions introduced by Dyson~\cite{dyson},
then used extensively
in the study of one-dimensional disordered systems~\cite{alea}.

\section{Statistics of the leader: one-time quantities}
\label{onetime}

\subsection{Invariant distribution}
\label{invar}

There is an invariant distribution
\beq
f_R=\lim_{m\to\infty}f_{R_m}
\eeq
associated with the random recursion~(\ref{rec}).
The latter implies a recursion between the probability
densities of $R_m$ and $R_{m+1}$:
\beq
f_{R_{m+1}}=(L+S)f_{R_m},
\label{rec_dens}
\eeq
where we have introduced the linear operators $L$ (leader) and $S$
(subleader),
acting on a function $f(x)$ defined for $0<x<1$ as
\beq
L f(x)=\int_0^{\min(1,x/(1-x))}\,\d uf(u),\quad
S f(x)=\int_x^{\min(1,x/(1-x))}\,\frac{\d u}{u}f(u).
\label{lsdef}
\eeq
The invariant distribution therefore obeys the fixed-point equation
\beq\
f_R=(S+L)f_R.
\label{fp}
\eeq

Defining the variable $V=1/R$, with density $f_V(v)$ for $v>1$,
eq.~(\ref{fp}) can be recast as the integral equation
\beq
v^2\,f_V(v)=\left\{\matrix{
\ds\int_1^v\d u\,u f_V(u)+1\hfill&(1<v<2),\cr\cr
\ds\int_{v-1}^v\d u\,u f_V(u)+\int_{v-1}^\infty\d uf_V(u)\quad&(v>2),\hfill
}\right.
\label{fpv}
\eeq
which simplifies to the differential equation
\beq
\frac{\d}{\d v}(v f_V(v))=f_V(v)+vf_V'(v)=\left\{\matrix{
0\hfill&(1<v<2),\cr
-f_V(v-1)\quad&(v>2).\hfill
}\right.
\label{fpvdiff}
\eeq
We readily obtain both the explicit result
\beq
f_V(v)=\frac{1}{v}\quad(1<v<2),
\label{fpone}
\eeq
and the integral relation
\beq
f_V(v)=\frac{1}{v}\int_{v-1}^\infty\d u\,f_V(u)\quad(v>2).
\label{fpformal}
\eeq
Transforming back to the variable $R$, we find
\beq\label{eq:fR}
f_R(x)=\frac{1}{x}\int_0^{\min(1,x/(1-x))}\,\d u f_R(u).
\eeq

Using~(\ref{fpone}) as an input,
the second line of~(\ref{fpvdiff}) can be solved iteratively.
We thus obtain more and more complex analytic expressions
for the invariant density $f_V(v)$ on the intervals delimited
by the integers.
We have $f_V(v)=(1-\ln(v-1))/v$ for $2<v<3$,
whereas the expression for $3<v<4$ involves the dilogarithm function
${\rm Li}_2(2-v)$.
The corresponding expressions in terms of the variable $R$ are
$f_R(R)=1/R$ for $1/2<R<1$ and $f_R(R)=[1+\ln(R/(1-R))]/R$ for $1/3<R<1/2$.
The invariant density $f_V(v)$ has weaker and weaker singularities
at integer values of $v$.
Using~(\ref{fpvdiff}) it is readily found that
the $n$--th derivative of $f_V(v)$ has a discontinuity at $v=n+1$
of the form
\beq
\D f_V^{(n)}(n+1)=f_V^{(n)}(n+1^+)-f_V^{(n)}(n+1^-)=\frac{(-1)^n}{(n+1)!}.
\eeq

\begin{figure}[!ht]
\begin{center}
\includegraphics[angle=90,width=.5\textwidth]{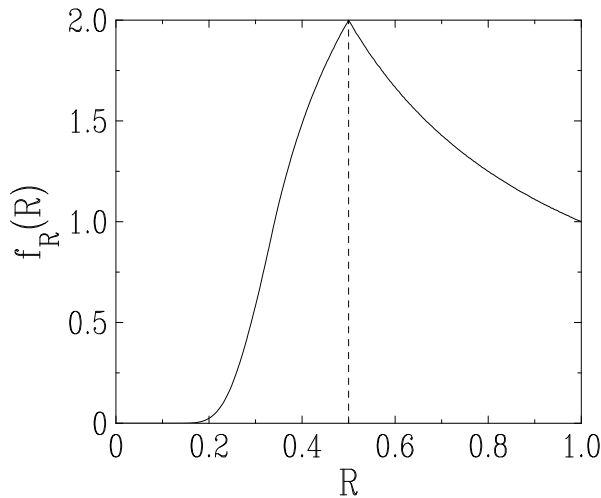}
\caption{
Plot of the invariant probability density $f_R(R)$.
The dashed line emphasizes the symmetric cusp at $R_\cusp=1/2$.}
\label{r}
\end{center}
\end{figure}

Figure~\ref{r} shows a plot of the invariant density $f_R(R)$,
obtained by a direct numerical simulation of the recursion~(\ref{rec}).
This procedure is more accurate than for example a numerical inversion
of the explicit form~(\ref{flap}) of the Laplace transform.
The leading singularity at $R=1/2$ is clearly visible as the maximum
of a symmetric cusp.
The above expressions for the invariant density indeed yield
$f'_R(R=1/2\pm0)=\mp4$.
The other singularities at $R=1/3$, $R=1/4$, etc., are not visible.

Another consequence of the differential equation~(\ref{fpvdiff}) is as follows.
Multiplying both sides of these equations by $v^p$,
where $p$ is any complex number, and integrating over $v$ gives
\beq
\mean{V^p}+\int_1^\infty\d v\,v^{p+1}f_V'(v)=-\mean{(V+1)^p}.
\eeq
An integration by parts using the initial value $f_V(1)=1$
coming from~(\ref{fpone}) leads to the identity
\beq
\mean{(V+1)^p}=1+p\mean{V^p},
\eeq
which holds for any complex $p$.
It translates into an identity for the
invariant distribution of the variable $R$
(up to a change of $p$ into its opposite)
\beq
\bigmean{\left(\frac{R}{1+R}\right)^p}=1-p\mean{R^p}.
\label{idenr}
\eeq

The explicit computation of the invariant distribution
is more easily performed by introducing the Laplace transform
\beq
\hat f_V(s)=\mean{\e^{-sV}}=\int_1^\infty\d v\, f_V(v)\,\e^{-sv}.
\label{hatdef}
\eeq
Eq.~(\ref{fpvdiff}) yields, using again the initial value $f_V(1)=1$,
\beq
s\,\frac{\d\hat f_V(s)}{\d s}=\e^{-s}(\hat f_V(s)-1).
\eeq
The solution of this equation reads
\beq
\hat f_V(s)=1-\e^{-E(s)}=1-s\,\e^{\euler-F(s)},
\label{flap}
\eeq
where we have introduced the functions
\beqa
E(s)&=&\int_s^\infty\d u\,\frac{\e^{-u}}{u}
=-{\rm Ei}(-s)=-\euler-\ln s+F(s),\nonumber\\
F(s)&=&\int_0^s\d u\,\frac{1-\e^{-u}}{u}
=\sum_{k\ge1}\frac{(-1)^{k+1}}{k\,k!}\,s^k
\eeqa
and Ei is the exponential integral function.
The second expression of the function $F(s)$
shows that it is an entire function of the complex variable $s$.
The moments of $V$ are readily obtained by expanding the second
expression of~(\ref{flap}) in powers of $s$.
They are rational multiples of $\e^\euler$:
\beq
\mean{V}=\e^\euler,\quad
\mean{V^2}=2\,\e^\euler,\quad
\mean{V^3}=\frac{9}{2}\,\e^\euler,\quad\hbox{etc.}
\eeq

Let us finally determine the behavior of $f_V(v)$ as $v\to\infty$,
or equivalently that of $f_R(x)$ as $x\to0$.
This analysis is conveniently done along the lines of~\cite{ln}.
Anticipating a fast decay in the regime under consideration, we set
\beq
f_V(v)\sim\e^{-\phi(v)},
\eeq
and approximate the integral in the right-hand side of~(\ref{fpformal}) by
\beq
\int_{v-1}^\infty\d u\,f_V(u)\approx\frac{\e^{-\phi(v-1)}}{\phi'(v-1)},
\eeq
obtaining thus
\beq
\phi(v)-\phi(v-1)\approx\ln v+\ln\phi'(v-1).
\eeq
Setting
\beq
\l=\ln v,\quad
\phi(v)=v\,a(\l),\quad
b(\l)=a(\l)+\frac{\d a(\l)}{\d\l},
\eeq
we obtain
\beq
b(\l)\approx\l+\ln b(\l).
\eeq
The latter equation is correct up to terms of relative order $1/v$,
i.e., exponentially small in $\l$.
It is therefore exact to all orders in $1/\l$.
Its solution yields the asymptotic expansion
\beq
a=\l+\mu+\frac{\mu}{\l}-\frac{(\mu-1)^2}{2\l^2}
+\frac{2\mu^3-9\mu^2+12\mu-13}{6\l^3}+\cdots,
\label{afact}
\eeq
with the shorthand notation $\mu=\ln\l-1=\ln(\ln v)-1$.
We thus obtain
\beq
f_V(v)\sim\exp\left\{-v\left(\ln v+\ln(\ln v)-1+\frac{\ln(\ln v)-1}{\ln v}
+\cdots\right)\right\}.
\label{ffact}
\eeq
The invariant density therefore decays faster than exponentially.
It can be said to decay {\it factorially},
as the leading behavior of~(\ref{ffact}) is identical to that of $1/v!$.

\subsection{Probability for a record node to be a leader}
\label{secomega}

The knowledge of the invariant distribution $f_R$ allows the determination of
one-time quantities in the late time regime of the process.
In particular, consider the asymptotic probability $\omega$ for a record node
to be a leader (for some period of time):
\beq
\omega=\lim_{m\to\infty}\omega_m=\lim_{m\to\infty}\prob(\LL_m)
=\lim_{m\to\infty}\prob\left(U_{m+1}<\frac{1}{1+R_m}\right).
\label{omegcalcul}
\eeq
Thus, using the identity~(\ref{idenr}) for $p=1$,
\beq
\omega=\bigmean{\frac{1}{1+R}}=\mean{R}=\int_0^\infty\d s\hat f_V(s).
\label{omeg}
\eeq
Finally, using~(\ref{flap}), we obtain
\beq
\omega=\int_0^\infty\d s\,\e^{-s-E(s)}
=\int_0^1\d x\,\exp\left(\int_0^x\frac{\d y}{\ln y}\right)
=0.624\,329\,988\dots
\label{omega}
\eeq

This number is known as the Golomb-Dickman constant~\cite{golombdickman}.
It first appeared in the framework of the decomposition
of an integer into its prime factors~\cite{dickman}, then in the study of the
longest cycle in a random permutation of order
$n$~\cite{golomb,goncharov,shepp}.
The connection of the present study to the statistics of cycles of permutations
will be given in section~\ref{connection1}.

Eq.~(\ref{omeg}) shows that $\omega$ is also the mean scaled maximal inter-record interval.

\subsection{Degree of the leader at a generic instant of time}
\label{secgeneric}

The invariant distribution $f_R$
associated with the recursion~(\ref{rec})
gives the probability distribution of the
fraction of nodes which are connected to the leader at a record time,
in the late-time regime.
We now solve the same question when the time of observation is a {\it generic}
late time.

Let an instant of time $n\gg1$ be given.
Eq.~(\ref{fund}) now reads
\beq
L(n)=\max(L_M,n-N_M),
\label{fund2}
\eeq
where $M\equiv M_n$ is the fluctuating number of records before time $n$,
and $L_M=\max(\D_2,\dots,\D_M)$.
Setting
\beq
R(n)=\frac{L(n)}{n},
\label{rndef}
\eeq
equation~(\ref{fund2}) leads to
\beq
R(n)=\max(R_M X_n,1-X_n),
\label{rec1}
\eeq
where the ratio $X_n$, defined in~(\ref{xyzdef}), is uniformly distributed
between 0 and 1 and independent of $R_M$ (see~(\ref{xyzlaw})).
The recursion~(\ref{rec1}) therefore maps $R_M$ onto $R(n)$
in exactly the same way as the recursion~(\ref{rec}) maps $R_m$ onto $R_{m+1}$.
As a consequence, the distribution of the ratio $R(n)$ at a generic late time
is also given by the invariant distribution $f_R$.
Stated otherwise,
the fraction of nodes which are connected to the leader at a record time
and at a generic time are identically distributed.\footnote{This property
extends in a straightforward way to the full degree statistics (see
section~\ref{connection2}).
The moments $Y^\p_m$ of any finite order $p$, introduced in~(\ref{ydef}),
are also identically distributed at a record time and at a generic time.}

\subsection{Probabilities for the current record node to be a leader}

Consider first the probability that the current record node at time $n$,
that is, the record node number $M_n$, is the leader at the current time $n$.
This event requires that $n-N_M>L_M$,
which implies $1-X_n>R_MX_n$, or finally $X_n<1/(1+R_M)$.
The variable~$R_M$ is asymptotically distributed according to the invariant
distribution~$f_R$, while $X_n$, uniformly distributed between~0 and 1,
is independent of $R_M$.
As a consequence, we find that the probability under consideration
is again equal to the Golomb-Dickman constant $\omega$.

Consider now the probability that the current record node at time $n$,
that is, the record node number $M_n$, is a leader (for some period of time).
This is a different quantity, larger than the previous one.
Recall that by ``a given node is a leader''
we mean that it leads for some duration of time.
It does not necessarily lead at time $n$,
but will surely lead at time $N_{M+1}$ (see~(\ref{leadineqs})).
Now the inequality to be satisfied is $N_{M+1}-N_M>L_M$, implying $Z_n>1+R_M$,
where the time ratio $Z_n$, defined in~(\ref{xyzdef}),
with probability density~(\ref{xyzlaw}), is independent of $R_M$.
The probability under consideration therefore reads
\beqa
\Omega=\int_0^1\d R\,f_R(R)\int_{1+R}^\infty\d z\,\frac{\ln z}{z^2}
=\bigmean{\frac{1+\ln(1+R)}{1+R}}
=0.914\,063\dots
\eeqa
As expected, we have $\Omega>\omega$.
This inequality has yet another interpretation.
The probability~$\Omega$ for {\it the current} record node to be a leader
is larger than the probability $\omega$ for {\it any} record node
to be a leader.
This is so because, for a fixed time $n$,
larger intervals $N_{M+1}-N_M$ have a higher probability to be sampled.
And the later the successor of a record node is born,
the higher its probability to be a leader.

\section{Statistics of the leader: two-time quantities}
\label{twotime}

This section is devoted to a study of two-time quantities in the process,
focusing our attention onto the main characteristics
of the reign of a leader, defined in section~\ref{process:discrete}, i.e.,
the difference $\nu$ of the labels of two successive leaders
(section~\ref{difflabel}),
and the time ratios $N_{m+\nu}/N_m$, $a_m/N_m$, $b_m/N_m$, $l_m/N_m$
(section~\ref{twora}).
All these quantities have limit distributions in the late-time regime.

\subsection{Statistics of the difference of the labels of two successive
leaders}
\label{difflabel}

Let us first consider the distribution of the random integer $\nu=1,2,\dots$,
which is the difference of the labels of two successive
leaders in the sequence of record times, $m$ and $m+\n$.
Hereafter we consider $m$ as being large enough,
so that the process is described by the continuum formalism.
We shift the record labels by~$m$,
so that the two leaders now have labels 0 and $\nu$.
We consistently denote the invariant density $f_R$ by $f_0$.
The probability for a record node to be a leader reads
\beq
\prob(\LL_0)=\omega.
\eeq
The distribution of $\n$ is encoded in the probabilities
\beqa
P_n=\prob(\nu=n)=\frac{1}{\omega}\,\prob(\LL_0\SS_1\dots\SS_{n-1}\LL_n),
\nonumber\\
Q_n=\prob(\nu\ge n)=\frac{1}{\omega}\,\prob(\LL_0\SS_1\dots\SS_{n-1}).
\label{pqdef}
\eeqa
The following relationships hold:
\beqa
&&P_n=Q_n-Q_{n+1},\quad Q_n=\sum_{m\ge n}P_m,\nonumber\\
&&\sum_{n\ge1}P_n=Q_1=1,\quad
\mean{\nu}=\sum_{n\ge1}n\,P_n=\sum_{n\ge1}Q_n.
\label{pqrel}
\eeqa

In order to compute the $P_n$ and $Q_n$ we consider the sequence of functions
$f_n(x)$ defined by
\beq
f_n=S^{n-1}Lf_0\quad(n\ge1),
\eeq
obtained as the result of the action, upon the invariant density $f_0$,
of the operator $L$,
followed by $n-1$ successive actions of the linear operator $S$.
We have
\beq
Q_n=\frac{1}{\omega}\int_0^1\d u\,f_n(u).
\label{qint}
\eeq
Similarly,
\beq
P_n=\frac{1}{\omega}\int_0^1\d u\,f_n(u)\,\frac{1}{1+u},
\label{pint}
\eeq
the factor $1/(1+u)$ being due to the last event $\LL_n$
in the definition~(\ref{pqdef}).

We have
\beq
\sum_{n\ge1}f_n=(1-S)^{-1}Lf_0=f_0
\label{fsum}
\eeq
since $(1-S)f_0=Lf_0$ by~(\ref{fp}), or $(1-S)^{-1}Lf_0=f_0$.
Eq.~(\ref{fsum}) has a simple interpretation:
the distribution of the variable~$R$ of the next leader
is identical to that of the variable $R$ of the current leader.
Inserting the sum rule~(\ref{fsum}) into~(\ref{qint}) yields,
using~(\ref{pqrel}),
\beq
\mean{\nu}=\sum_{n\ge1}Q_n=\frac{1}{\omega}.
\label{nuave}
\eeq
This result, too, has a simple interpretation:
the mean distance $\mean{\nu}$ between two consecutive leaders
along the sequence of record times
is the inverse of the probability~$\omega$ for a given record to be a leader,
or the inverse record breaking rate for the sequence of inter-record intervals.

We now perform the computation of the cumulative probabilities $Q_n$.
Let
\beq
f_1=Lf_0,
\label{f1def}
\eeq
i.e., using~(\ref{lsdef}),
\beq
f_1(x)=
\left\{\matrix{
1\hfill&\mbox{for }1/2\le x\le1,\hfill\cr
\ds\int_0^{x/(1-x)}\d u\,f_0(u)\quad&\mbox{for }0\le x\le1/2,\hfill
}\right.
\label{f1bin}
\eeq
or finally, using (\ref{eq:fR}),
\beq\label{eq:f1}
f_1(x)=x f_0(x).
\eeq
The function $f_1(x)$ is the initial condition for the recursion
\beq
f_{n+1}(x)=S f_n(x)=\int_x^{\min(1,x/(1-x))}\,\frac{\d u}{u}f_n(u).
\eeq
We make the change of variable $y=1/x$ and define $g_n(y)=f_n(x)$.
These functions obey the recursion
\beq
g_{n+1}(y)=\int_{\max(1,y-1)}^y\,\frac{\d v}{v}g_n(v),
\eeq
which assumes a simpler form in terms of Laplace transforms:
\beq
\hat g_n(s)=-\frac{\d}{\d s}\left(\frac{s}{1-\e^{-s}}\hat g_{n+1}(s)\right).
\label{gns}
\eeq
Consider the generating function of the $\hat g_n(s)$,
\beq
G(x,s)=\sum_{n\ge2}\hat g_n(s) x^{n-1}.
\eeq
From~(\ref{gns}) we obtain the differential equation
\beq
x(G(x,s)+\hat g_1(s))=-\frac{\d}{\d s}\left(\frac{s}{1-\e^{-s}}\,G(x,s)\right),
\eeq
whose solution reads
\beq
G(x,s)=x\,\frac{1-\e^{-s}}{s}\,\e^{-xF(s)}\int_s^\infty
\d t\,\e^{xF(t)}\,\hat g_1(t).
\label{solution}
\eeq
Furthermore, from (\ref{eq:f1}) it is easily found that
\beq
\hat g_1(s)=-\frac{\d}{\d s}\hat f_V(s)=\e^{\euler-s-F(s)}.
\label{solinit}
\eeq

Coming back to the $Q_n$, we have
\beq
Q_n=\frac{1}{\omega}\int_0^1\d u f_n(u)
=\frac{1}{\omega}\int_1^\infty\d v\,\frac{g_n(v)}{v^2}
=\frac{1}{\omega}\int_0^\infty\d s\,s\,\hat g_n(s).
\eeq
Introducing the generating function of the $Q_n$,
\beq
\Sigma(x)=\sum_{n\ge1}Q_nx^{n-1},
\eeq
we obtain
\beq
\Sigma(x)=1+\frac{1}{\omega}\int_0^\infty\d s\,s\,G(x,s).
\eeq
The expressions~(\ref{solution}) and~(\ref{solinit}) yield the explicit result
\beq
\Sigma(x)=\frac{\e^\euler}{\omega}\int_0^\infty\d s\,\e^{-xF(s)}
\int_s^\infty\d t\,\e^{-t+(x-1)F(t)}.
\label{solex}
\eeq

Several results of interest can be derived from this exact expression.
First of all, taking the derivative at $x=0$, we obtain
\beq
Q_2=\Sigma'(0)
=\frac{1}{\omega}\int_0^\infty\frac{\d t}{t}(t-1+\e^{-t})\,\e^{-t-E(t)}.
\eeq
The integral can be evaluated to yield
\beq
Q_2=\frac{2\omega-1}{\omega},\quad
\hbox{i.e.,}\quad P_1=\frac{1-\omega}{\omega}.
\eeq
This can alternatively be obtained using (\ref{pint}) and (\ref{eq:f1}).
The probability that two successive records are leaders
therefore reads $\prob(\LL_0\LL_1)=\omega P_1=1-\omega=0.375\,670\dots$
This value is close to what it would be
in the absence of any correlation, namely $\omega^2=0.389\,788\dots$

The moments of $\nu$ can be derived by expanding~(\ref{solex}) around $x=1$.
We thus recover the result~(\ref{nuave}) for $\mean{\nu}$,
as $\Sigma(1)=1/\omega$.
For the second moment $\mean{\nu^2}$, we find
\beqa
\mean{\nu^2}&=&\sum_{n\ge1}n^2P_n=\sum_{n\ge1}(2n-1)Q_n
=\frac{1}{\omega}+2\Sigma'(1)
\nonumber\\
&=&\frac{1}{\omega}\left[1+2\int_0^\infty\frac{\d s}{s}\,\e^{-E(s)}
\int_s^\infty\frac{\d t}{t}\,\e^{-t}(1-\e^{-t})\right]
\nonumber\\
&=&3.383\,695\dots
\eeqa

The expression~(\ref{solex}) also shows that $\Sigma(x)$
has no singularity at any (finite) value of $x$.
In other words, it is an entire function of the complex variable $x$.
As a consequence,
the $Q_n$ fall off at large $n$ faster than exponentially.
This means that $Q_{n+1}/Q_n\to0$, so that $P_n\approx Q_n$.
From a quantitative viewpoint,
the fall-off of the $Q_n$ (i.e., of the $P_n$)
can be derived from the asymptotic
behavior of $\Sigma(x)$ as $\Re x\to+\infty$.
In this regime, the double integral in~(\ref{solex})
is dominated by small values of $s$, where $F(s)\approx s$,
and large values of $t$, where $F(t)\approx\ln t+\euler$.
We thus obtain
\beq
\Sigma(x)\approx\frac{\e^{\euler x}}{\omega x}\,\Gamma(x)
\approx\frac{1}{\omega}\,\sqrt{\frac{2\pi}{x^3}}\,\e^{x(\ln x+\euler-1)}
\quad(\Re x\to+\infty).
\eeq
The contour integral representation of $Q_n$,
\beq
Q_n=\oint\frac{\d x}{2\pi\i}\,\frac{\Sigma(x)}{x^n},
\eeq
can then be evaluated by the saddle-point method.
Parametrizing the saddle-point as $x_c=n/a$,
we are left with the estimate
\beq
P_n\approx Q_n\approx\frac{a}{\omega n\sqrt{a+1}}\,\e^{-n(a-\euler-1+1/a)},
\label{pqasy}
\eeq
where $a$ obeys the implicit equation
\beq
a=\ln(n/a)+\euler.
\label{aimp}
\eeq
Setting $\lambda=\ln n+\euler$ and $\mu=\ln\lambda=\ln(\ln n+\euler)$,
we obtain the asymptotic expansions
\beq
a=\lambda-\mu+\frac{\mu}{\lambda}+\frac{\mu(\mu-2)}{2\lambda^2}+\cdots
\eeq
and
\beq
P_n\approx Q_n
\sim\exp\left\{-n\left(\ln
n-\mu-1+\frac{\mu+1}{\lambda}+\frac{\mu^2}{2\lambda^2}
+\cdots\right)\right\}.
\label{qfact}
\eeq
The above expansions are very similar to~(\ref{afact}) and~(\ref{ffact}).
The $P_n$ can again be said to decay {\it factorially},
as their leading behavior coincides with that of $1/n!$.

\begin{figure}[!ht]
\begin{center}
\includegraphics[angle=90,width=.5\textwidth]{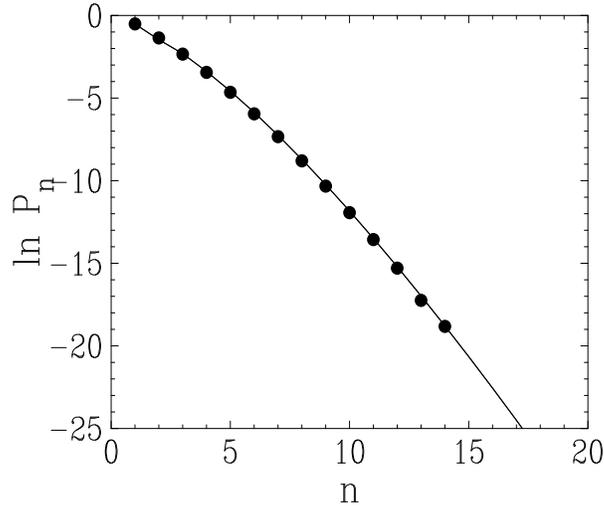}
\caption{
Probability distribution of the difference of labels $\nu$
between two successive leaders.
The logarithm of $P_n=\prob(\nu=n)$ is plotted against the integer $n$.
Symbols: data obtained by a direct numerical simulation
of the random recursion~(\ref{rec}).
Line: value of the asymptotic estimate~(\ref{pqasy}),
obtained by means of a numerical solution
of the implicit equation~(\ref{aimp}).}
\label{pn}
\end{center}
\end{figure}

The accuracy of the estimate~(\ref{pqasy}) is demonstrated in Figure~\ref{pn},
showing a surprisingly good agreement between this prediction
and numerical data for the pro\-ba\-bi\-li\-ty distribution $P_n$,
obtained by a direct numerical simulation of the recursion~(\ref{rec}).

\subsection{Statistics of the lengths of time associated with the reign of a
leader}
\label{twora}

We now turn to the statistics of the time ratios
\beq
\rho_m=\frac{N_{m+\nu}}{N_m},\quad
\alpha_m=\frac{a_m}{N_m},\quad
\beta_m=\frac{b_m}{N_m},\quad
\lambda_m=\frac{l_m}{N_m},
\eeq
where $N_m$ and $N_{m+\nu}$ are the birth times of the current leader
and of the next one,
and where the beginning and ending times, $a_m$ and $b_m$,
and the duration of the reign~$l_m$ of the current leader
are defined in~(\ref{cdef}), (\ref{adef}) and~(\ref{tdef}).

We again shift the record labels by~$m$,
so that the two leaders under consideration have labels 0 and $\nu$,
whereas the intermediate record nodes, labeled $m=1,\dots,\nu-1$,
are not leaders.
We have therefore, using~(\ref{rec})
\beqa
&&R_1=1-U_1,\nonumber\\
&&R_2=(1-U_1)U_2,\quad\dots,\nonumber\\
&&R_\nu=(1-U_1)U_2\dots U_\nu.
\eeqa
The above time ratios can therefore be expressed
in terms of the variables $R_0$, $R_1$ and~$R_\nu$~as
\beqa
\alpha_0&=&1+R_0,
\label{tic}
\\
\rho_0&=&\frac{R_1}{(1-R_1)R_\nu},\quad
\beta_0=\frac{R_1(1+R_\nu)}{(1-R_1)R_\nu}=(1+R_\nu)\rho_0,\quad
\label{tira}
\\
\lambda_0&=&\beta_0-\alpha_0.
\label{tit}
\eeqa
These time ratios have well-defined limit distributions
$f_\alpha$, $f_\rho$, $f_\beta$, $f_\lambda$ in the late-time regime.
This reflects the temporal self-similarity of the process.

\begin{figure}[!ht]
\begin{center}
\includegraphics[angle=90,width=.5\textwidth]{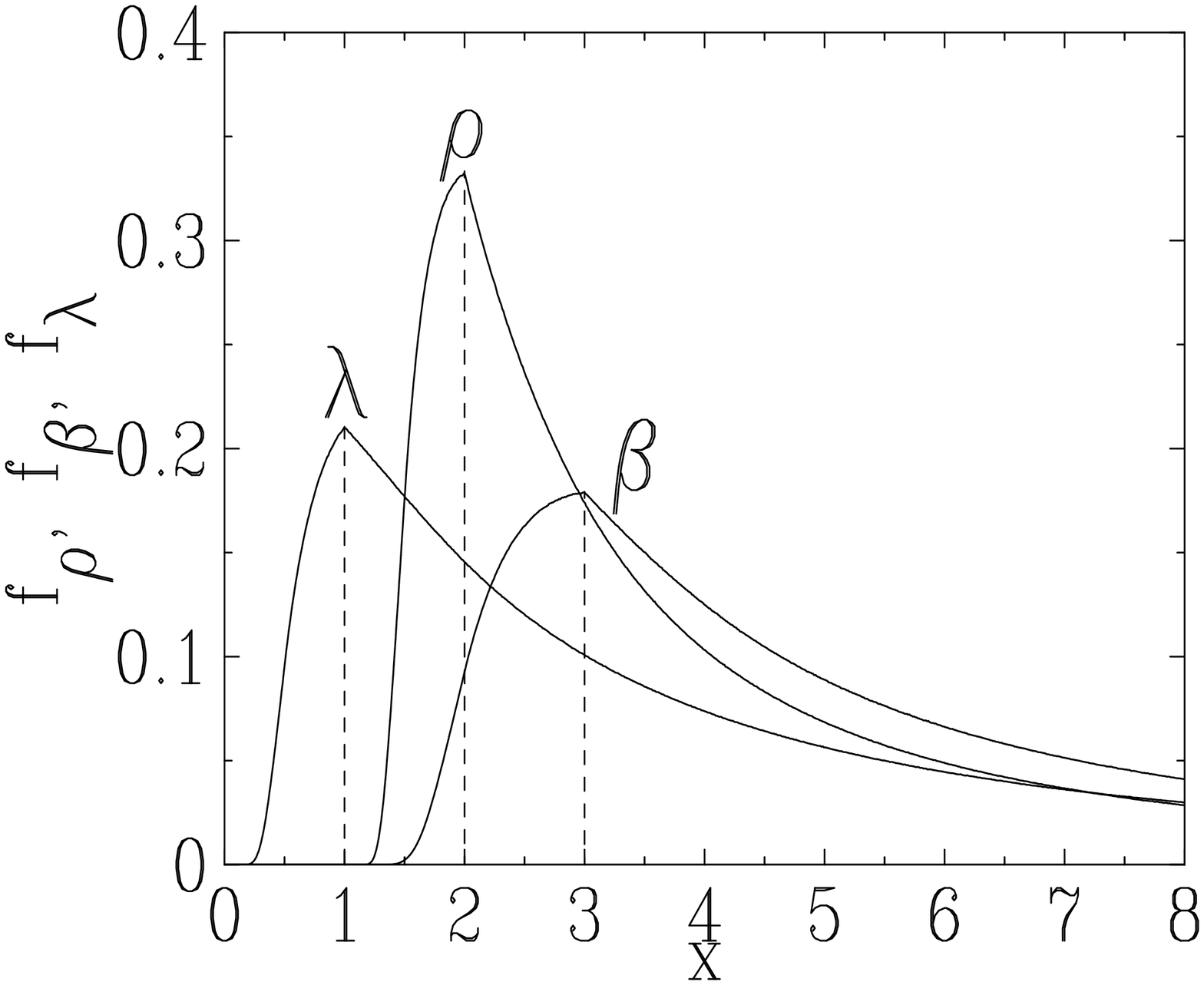}

{\hskip 3pt}\includegraphics[angle=90,width=.52\textwidth]{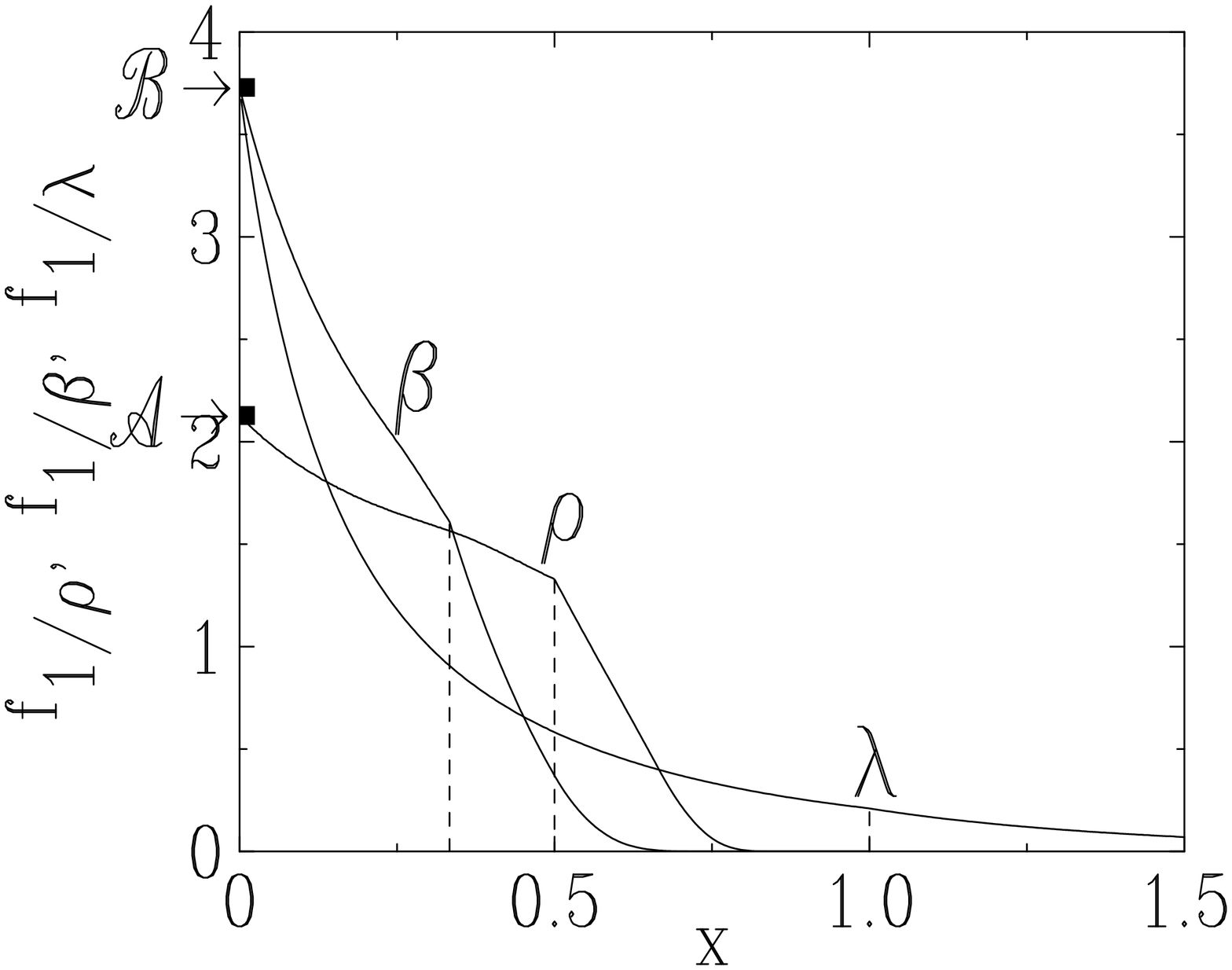}
\caption{
Top: plot of the probability distributions $f_\rho$, $f_\beta$, $f_\lambda$
of the time ratios introduced in~(\ref{tira}),~(\ref{tit}).
Bottom: plot of the probability distributions of their inverses.
Symbols: exact values $\A$ and $\B$ of the densities at the origin
(see~(\ref{abrel})--(\ref{abnum})).
The dashed lines emphasize the cusps
at $\rho_\cusp=2$, $\beta_\cusp=3$ and $\lambda_\cusp=1$.}
\label{rat}
\end{center}
\end{figure}

Figure~\ref{rat} shows plots of the probability distributions
$f_\rho$, $f_\beta$, $f_\lambda$ of the three
time ratios $\rho_0$, $\beta_0$ and $\lambda_0$,
and of the distributions $f_{1/\rho}$, $f_{1/\alpha}$, $f_{1/\lambda}$ of their
inverses.
The data have been obtained by a direct simulation of the recursion~(\ref{rec}).
The plots emphasize the following characteristics.
The three distributions $f_\rho$, $f_\beta$, $f_\lambda$
share with the invariant distribution $f_R$
the property that their maxima correspond to cusps.
The values at which these cusps occur,
namely $\rho_\cusp=2$, $\beta_\cusp=3$ and $\lambda_\cusp=1$,
are readily obtained by replacing in~(\ref{tira}),~(\ref{tit})
both variables $R_1$ and $R_\nu$ by $R_\cusp=1/2$.
The lower plot in Figure~\ref{rat} demonstrates that
the distributions of the inverse time ratios have well-defined limits at zero:
\beq
f_{1/\rho}(0)=\A,\quad
f_{1/\beta}(0)=f_{1/\lambda}(0)=\B.
\eeq
In other words, the distributions of these three time ratios
fall off as $1/x^2$:
\beq
f_{\rho}(x)\approx\frac{\A}{x^2},\quad
f_\beta(x)\approx f_\lambda(x)\approx\frac{\B}{x^2}.
\eeq
The exact values of the amplitudes $\A$ and $\B$ will be evaluated below
(see~(\ref{abrel})--(\ref{abnum})).
The fact that $f_\beta$ and $f_\lambda$ share the same fall-off amplitude $\B$
is simply due to the fact that the difference $\beta_0-\lambda_0=\alpha_0$
is bounded in the range $1\le\alpha_0\le 2$~(see~(\ref{tic})).

The actual evaluation of the joint distribution
of the variables $R_1$ and $R_\nu$ is now performed.
The starting point is to consider the probability
\beqa
P(r_1,r)&=\prob(R_1<r_1,R_\nu<r,\LL_0),\nonumber\\
&=\sum_{n\geq1}\prob(R_1<r_1,R_n<r,\LL_0\,,\n=n),\nonumber\\
&=\sum_{n\geq1}\prob(R_1<r_1,R_n<r,\LL_0\,\SS_1\dots\SS_{n-1}\,\LL_n).
\eeqa
Note that $P(1,1)=\prob(\LL_0)=\omega$.
The normalized joint density that we are looking for is thus
\beq
f_{R_1,R_\n}(r_1,r)
=\frac{1}{\omega}
\frac{\partial}{\partial r_1}\frac{\partial}{\partial r}P(r_1,r).
\eeq

As in section~\ref{difflabel},
we start from the invariant distribution $f_0$.
Acting on it with the operator $L$ yields $f_1$ (see~(\ref{f1def})).
We then fix the value of $R_1$ to $r_1$, thus obtaining the density
$(1/\omega)f_1(r_1)\delta(r-r_1)$.
We then define
\beq
\f_1(r_1,r)=\delta(r-r_1),
\eeq
and introduce the functions
\beqa
\f_n(r_1,r)&=&S^{n-1}\f_1(r_1,r),\nonumber\\
\f(r_1,r)&=&\sum_{n\geq1}\f_n(r_1,r)=(1-S)^{-1}\f_1(r_1,r),
\eeqa
where the operator $S$ acts on the variable $r$.

We are thus left with the result
\beq
f_{R_1,R_\n}(r_1,r)=
\frac{1}{\omega}\,f_1(r_1)\f(r_1,r)\,\frac{1}{1+r},
\label{jointres}
\eeq
where the factor $1/(1+r)$ comes from the last event $\LL_n$.
We now make the change of variable $y=1/r$, $y_1=1/r_1$,
and define $\g_n(y_1,y)=\f_n(r_1,r)$ and $\g(y_1,y)=\f(r_1,r)$.
The Laplace transform of $\g_1(y_1,y)$ is $\hat\g_1(y_1,s)=y_1^2\,\e^{-sy_1}$.
Therefore, in analogy with~(\ref{solution}),
the Laplace transform of $\g(y_1,y)$ reads
\beq
\hat\g(y_1,s)=
y_1^2\left(\e^{-sy_1}+\frac{1-\e^{-s}}{s}\,\e^{-F(s)}\int_s^\infty
\d t\,\e^{F(t)-ty_1}\right).
\label{lapgg}
\eeq

The expression~(\ref{jointres}) of the normalized joint
probability density of the variables~$R_1$ and $R_\nu$
therefore contains two non-trivial factors:
$f_1(r_1)=g_1(y_1)$,
whose Laplace transform with respect to $y_1$ is given by~(\ref{solinit}),
and $\f(r_1,r)=\g(y_1,y)$,
whose Laplace transform with respect to $y$ is given by~(\ref{lapgg}).
The knowledge of this distribution allows one, at least in principle,
to compute the distribution of the time ratios $\rho_0$, $\lambda_0$, $\beta_0$
and of similar quantities.
Analytical expressions thus obtained are however very cumbersome,
and therefore of little use, either theoretically or practically,
except for some simple examples,
such as the amplitudes $\A$ and~$\B$, which will now be evaluated explicitly.

Consider first the amplitude $\A$.
The probability density $f_{1/\rho}$ can be expressed~as
\beqa
f_{1/\rho}(x)
&=&\int_0^1\d r_1\int_0^1\d r\,f_{R_1,R_\n}(r_1,r)
\;\delta\!\left(x-\frac{(1-r_1)r}{r_1}\right)\\
&=&\int_0^1\d r\,
\frac{r}{(r+x)^2}\,
f_{R_1,R_\n}\left(\frac{r}{r+x},r\right).
\eeqa
This expression simplifies in the $x\to0$ limit to yield
\beq
\A=f_{1/\rho}(0)=\int_0^1\d r\,f_{R_1,R_\n}(1,r)\,\frac{1}{r}
=\frac{1}{\omega}\,\bigmean{\frac{1}{R_\nu}}_{R_1=1}.
\eeq
We obtain similarly
\beq
\B=f_{1/a}(0)=\int_0^1\d r\,f_{R_1,R_\n}(1,r)\,\frac{r+1}{r}
=\frac{1}{\omega}\,\bigmean{1+\frac{1}{R_\nu}}_{R_1=1},
\eeq
hence the relation
\beq
\B=\A+\frac{1}{\omega}.
\label{abrel}
\eeq
The explicit form of $\B$ is
\beq
\B=\frac{1}{\omega}\int_0^1\frac{\d r}{r}\,\f(1,r)
=\frac{1}{\omega}\int_0^\infty\d s\,\hat\g(1,s)
=\frac{1}{\omega}\int_0^\infty\d t\,\e^{-t+F(t)},
\eeq
where the double integral resulting from the insertion
of the expression~(\ref{lapgg}) of $\hat\g(1,s)$ has been simplified
by an integration by parts.
We thus obtain the numerical values
\beq
\A=2.127\,451\dots,\quad\B=3.729\,168\dots
\label{abnum}
\eeq

\section{Connections to other fields}
\label{connection}

\subsection{Records and cycles of permutations}
\label{connection1}

As noted above, the asymptotic probability for a record node to be a leader is
equal to the Golomb-Dickman constant, a number which appears in problems of
combinatorical nature.
It is for example the limit, when $n\to\infty$, of $\langle\Lambda_n\rangle/n$,
where $\Lambda_n$ is the length of the longest cycle in a random permutation of
order $n$~\cite{golomb,goncharov,shepp}.
This number also appears in the framework of the decomposition
of an integer into its prime factors~\cite{dickman}.

Furthermore, as shown by Goncharov~\cite{goncharov} and Shepp and
Lloyd~\cite{shepp}, $\Lambda_n/n\to\cal{R}$, where the distribution of
the limiting random variable $\cal{R}$
coincides with the invariant distribution $f_R$ found in the present work.
The identity of the asymptotic probability for a record node to be a leader
and of the Golomb-Dickman constant is just the identity of
$\langle R\rangle$ (see eq.~(\ref{omeg})) and $\langle\cal{R}\rangle$.

We now explain the origin of the coincidences between features of the
statistics of records and the cycle structure of permutations.
The existence of connections between the two fields is well
known~\cite{ren,gold}.
In particular, $M_n$, the number of records up to time~$n$,
has the same distribution as the number of cycles $C_n$ in a random permutation
of order~$n$~\cite{ren}, as mentioned in section~\ref{records:discrete}.
There is actually a deeper relationship between
the sequence of record times on the one hand,
and the cycles of a random permutation on the other hand, which is due to the
fact that the latter can be generated
by the same set of indicator variables as the former.

For records, these variables are the $I_i$
defined in section~\ref{records:discrete}.
For cycles of permutations the construction is due to Feller~\cite{feller}, as
we now recall.
A permutation of order $n$ is constructed by a succession of $n$ decisions.
Let $a_1$, $a_2$, $\ldots$, $a_n$ be the $n$ letters of the permutation.
The position of $a_1$ is first chosen, with $n$ possibilities: $1\to i$.
Then the position of $a_i$ is chosen, with $n-1$ possibilities: $i\to j$, and
so on, until a cycle is formed.
For example~\cite{feller}, with $n=8$, choosing
\beqa
1\to3\to4\to1
\nonumber\\
2\to5\to6\to8\to2
\nonumber\\
7\to7,
\eeqa
generates the permutation $a_4\,a_8\,a_1\,a_3\,a_2\,a_5\,a_7\,a_6$.
Define the indicator variable $J_k$, equal to 1 if a cycle is formed at the
$k-$th step, else to 0.
For this example we have $J_3=1$, $J_7=1$, $J_8=1$.
Clearly, in general,
\beq
\prob(J_k=1)=\frac{1}{n-k+1},
\eeq
and the $J_k$ are independent.
Here the sequence of $J_k$ reads $0\,0\,1\,0\,0\,0\,1\,1$,
with $\prob(J_3=1)=1/6$, $\prob(J_7=1)=1/2$, $\prob(J_8=1)=1$.
Reverting the sequence of steps, we obtain $1\,1\,0\,0\,0\,1\,0\,0$ for the new
indicator variable, that we denote by $I_i$ say, such that $\prob(I_1=1)=1$,
$\prob(I_2=1)=1/2$, $\prob(I_6=1)=1/6$.
One recognizes the construction of a sequence of records.

\subsection{Random breaking of an interval}
\label{connection2}

We have been, up to now, mainly interested in the statistics of leaders.
We now address the full statistics of node degrees,
within the continuum approach.
Consider a fixed late record time $N_m$.
The degrees of the earlier records read
$K_{j-1}(N_m)=N_{j}-N_{j-1}=\Delta_{j}$ for $j=2,\dots,m$.
These degrees sum up to
$N_m$\footnote{The sum is actually equal to $N_m-1$, but we neglect the
correction in the continuum limit.},
each of them representing a finite fluctuating
fraction of the total, or ``weight'' $\Delta_{j}/{N_m}$.
We have
\beqa
&&\frac{\Delta_{m}}{N_m}=1-U_m,\nonumber\\
&&\frac{\Delta_{m-1}}{N_m}=U_m(1-U_{m-1}),\nonumber\\
&&\frac{\Delta_{m-2}}{N_m}=U_m U_{m-1}(1-U_{m-2}),\quad\dots
\eeqa
Relabeling the indices $j\to k=m-j+1$, and denoting the weights by $W_k$, we
have
\beqa
&&W_{1}=1-U_1,\nonumber\\
&&W_{2}=U_1(1-U_{2}),\nonumber\\
&&W_{3}=U_1U_2(1-U_{3}),\quad\dots
\eeqa
We recognize the sequence of weights obtained by randomly breaking an interval
of unit length into two pieces, and iterating the process~\cite{df}.

A convenient tool to investigate this kind of fluctuating weights
consists in introducing the reduced moments
\beq
Y^\p_m=\sum_{j=2}^{m}\left(\frac{\Delta_{j}}{N_m}\right)^p
=\sum_{k=1}^{m-1}W_k^p,
\label{ydef}
\eeq
where the order $p=1,2,\dots$ is any integer~\cite{df}.
These moments obey the random recursion
\beq
Y^\p_{m+1}=U_{m+1}^pY^\p_m+(1-U_{m+1})^p.
\label{recp}
\eeq
The latter can be viewed as a generalization
of the recursion~(\ref{rec}) to any finite integer order $p$.
Eq.~(\ref{rec}) is formally recovered in the $p\to\infty$ limit,
where the sum in~(\ref{ydef}) is dominated by its largest term,
that is by the contribution of the leader.
The $Y^\p_m$, which keep fluctuating in the late-time regime,
have non-trivial limit distributions,
invariant under the dynamical system~(\ref{recp}).
A plot of $Y^{(2)}_m$, obtained by iterating (\ref{recp})
numerically, as well a plot of the invariant measure $f_R$ of the largest
weight can be found in~\cite{df}.
In the framework of the random breaking of an interval,
$\omega$ is either the probability that the first weight $W_1$ be the largest,
or the mean maximal weight.

\subsection{Relation to the Kesten variable}
\label{connection3}

It turns out that the invariant distribution of the random variable $R$
can be worked out more generally for a one-parameter family
of problems containing the above as a special case.
Consider the recursion~(\ref{rec}) where the i.i.d.~random variables $U_{m+1}$
have an arbitrary distribution between 0 and 1, with density $\rho(u)$.
The invariant distribution still obeys
the fixed-point equation~(\ref{fp}), with the definitions
\beqa
S f(x)
&=&\int_x^{\min(1,x/(1-x))}\,\frac{\d u}{u}\,\rho\!\left(\frac{x}{u}\right)
\,f(u),
\nonumber\\
L f(x)
&=&\rho(1-x)\int_0^{\min(1,x/(1-x))}\,\d uf(u).
\eeqa
The integral equation thus obtained cannot be solved in closed form in general.
It nevertheless leads to a differential equation
similar to~(\ref{fpvdiff}), and is therefore solvable,
whenever the density of the variables $U_{m+1}$ is a power law.
More precisely, if
\beq
\rho(u)=bu^{b-1},
\eeq
where $b$ is an arbitrary positive parameter, one has
\beqa
\frac{\d}{\d v}\left(\frac{v^b}{(v-1)^{b-1}}f_V(v)\right)
&=&\frac{v^{b-1}}{(v-1)^b}\left[(v-b)f_V(v)+v(v-1)f_V'(v)\right]
\nonumber\\
&=&\left\{\matrix{
0\hfill&(1<v<2),\cr
-bf_V(v-1)\quad&(v>2).\hfill
}\right.
\label{bvdiff}
\eeqa
Consider the modified Laplace transform
\beq
\tilde f_V(s)=\mean{V^b\,\e^{-sV}}=\int_1^\infty\d v\,v^b\,\e^{-sv}\,f_V(v).
\eeq
This quantity obeys the differential equation
\beq
s\,\frac{\d\tilde f_V(s)}{\d s}=-(s+b(1-\e^{-s}))\tilde f_V(s),
\eeq
whose normalized solution reads
\beq
\tilde f_V(s)=\frac{\Gamma(b+1)}{s^b}\,\e^{-s-bE(s)}
=\Gamma(b+1)\,\e^{b\euler}\,\e^{-s-bF(s)}.
\label{ftilde}
\eeq
The probability $\omega$ (see eq.~(\ref{omegcalcul})) generalizes to
\beq
\omega(b)=\lim_{m\to\infty}\prob(1-U_{m+1}>U_{m+1}R_m)
=\bigmean{\left(\frac{1}{1+R}\right)^b},
\eeq
that is
\beq
\omega(b)=\bigmean{\left(\frac{V}{V+1}\right)^b}
=\frac{1}{\Gamma(b)}\int_0^\infty\d s\,s^{b-1}\,\e^{-s}\,\tilde f_V(s),
\eeq
i.e., finally
\beq
\omega(b)=\int_0^\infty\d s\,\e^{-s-bE(s)}.
\label{wb}
\eeq

The original problem with a uniform distribution of the variables $U_{m+1}$
is recovered by setting $b=1$ in the above results.
One has indeed $\tilde f_V(s)=-\d\hat f_V(s)/\d s$,
and an integration by parts shows that~(\ref{ftilde})
is equivalent to~(\ref{flap}).
Finally, the expression~(\ref{wb}) for $\omega(1)$
coincides with the expression~(\ref{omega}) of the Golomb-Dickman constant.

It can be checked that $\omega(b)=\mean{R}$ for all values of $b$.
This identity generalizes~(\ref{omeg}).
However, $\lim_{m\to\infty}\prob(1-U_{m+1}>U_{m+1}R_m)$ and $\mean{R}$ are not
equal for an arbitrary distribution $\rho(u)$.

The above family of exactly solvable invariant densities
is in correspondence with the following problem.
Consider the Kesten variable~\cite{kesten,calan}, defined as
\beq
Z=1+x_1+x_1x_2+x_1x_2x_3+\cdots,
\label{defkes}
\eeq
where the $x_m$ are i.i.d.~positive random variables
with probability density $\rho^\kes(x)$.
If this distribution is such that $\mean{\ln x}<0$,
the sum in~(\ref{defkes}) is convergent
and $Z$ has a well-defined
probability density $f_Z^\kes(z)$, solution of
the integral equation
\beq
f_Z^\kes(z)=\int_0^\infty\frac{\d x}{x}\,\rho^\kes(x)\,
f_Z^\kes\!\left(\frac{z-1}{x}\right),
\eeq
This equation again cannot be solved in closed form in general.
It is however known~\cite{calan,vervaat} that the problem can be solved
whenever the density of the variables $x_m$ is a power law
on an interval $[0,a]$.
In the marginal situation ($a=1$)
where the $x_m$ are between 0 and~1, with density
\beq
\rho^\kes(x)=bx^{b-1},
\eeq
where $b$ is again an arbitrary positive parameter, the Laplace transform
$\hat f_Z^\kes(s)=\mean{\e^{-sZ}}$ has the closed-form expression
\beq
\hat f_Z^\kes(s)=\e^{-s-bF(s)}.
\label{hatbkes}
\eeq
The similarity between the two problems is now patent by
comparing~(\ref{ftilde}) and~(\ref{hatbkes}).
The probability densities of the variable $V=1/R$ and of the Kesten variable
$Z$ are related to each other by the equation
\beq
x^b\,f_V(x)=\Gamma(b+1)\,\e^{b\euler}\,f_Z^\kes(x).
\eeq

\section{Conclusion}
\label{discussion}

The main goal of this work has been to put forward
the {\it Record-driven growth process}.
This ballistic growth model entirely based on the record process
has been met as the zero-temperature limit of a class
of network growth models with preferential attachment.
Its simplicity and its minimality however suggest
that the RD growth process might be relevant to a wider class of situations,
besides the realm of complex networks.
The main emphasis has been put on the interplay between
records (i.e., nodes endowed with the best intrinsic qualities)
and leaders (i.e., nodes whose degrees are the largest).
The RD growth process provides a natural playground
where subtle questions related to the statistics of leaders and of lead changes
can be addressed in a quantitative way.

The RD growth process inherits from the record process
some relationships with combinatorical problems related to permutations.
Relationships with fragmentation models
and with one-dimensional disordered systems have also been underlined.
A key feature of the RD growth process
is its temporal self-similarity in the late-time regime,
inherited from the underlying record process,
which manifests itself in that various time ratios
have non-trivial limiting distributions in the regime of late times.
This regime is also characterized by the very fast fall-off of temporal correlations.

Let us point out the recent work~\cite{maj}, which addresses the statistics
of records for the successive positions of a random walk.
The mean longest duration of a record scales with the number of steps,  
the ratio defining a non-trivial constant $0.626\,508\dots$
This parallels the scaling of the mean maximal inter-record interval  
of the present work, resulting in the occurrence of the Golomb-Dickman  
constant $0.624\,329\dots$

To close up, it is worth looking back to our starting point,
namely the growing networks with preferential attachment
considered in the Introduction.
The self-similar growth regime of the RD process
can be shown to be unstable against thermal fluctuations.
This regime crosses over to a complete freeze-out at a time scale
which usually diverges at a power-law at low temperature, as $\tau\sim T^{-b}$.
The model-dependent exponent $b$ can be evaluated
by generalizing the line of thought of the recent work~\cite{fb},
whether or not the model has a finite-temperature condensation transition.

\subsection*{Acknowledgments}

It is a pleasure for us to thank Ginestra Bianconi for stimulating discussions.


\section*{References}

\begin{thebibliography}{99}

\bibitem{ba} Barab\'asi A L and Albert R, 1999 {\it Science} {\bf 286} 509

\bibitem{abrmp} Albert R and Barab\'asi A L, 2002 {\it Rev. Mod. Phys.} {\bf 74} 47

\bibitem{doro} Dorogovtsev S N and Mendes J F F, 2003 {\it Evolution of Networks} (Oxford: Oxford University Press)

\bibitem{blmch} Boccaletti S, Latora V, Moreno Y, Chavez M and Hwang D U, 2006 {\it Phys. Rep.} {\bf 424} 175

\bibitem{cup} Barrat A, Barth\'elemy M and Vespignani A, 2008 {\it Dynamical Processes on Complex Networks} (Cambridge: Cambridge University Press)

\bibitem{bb} Bianconi G and Barab\'asi A L, 2001 {\it Europhys. Lett.} {\bf 54}, 439

\nonum Bianconi G and Barab\'asi A L, 2001 {\it Phys. Rev. Lett.} {\bf 86} 5632

\bibitem{ren} Renyi A, 1962 Proceedings Coll. Combinatorial Methods in Probability Theory (Math. Inst. Aarhus Univ., Aarhus, Denmark)

\bibitem{glick} Glick N, 1978 {\it Amer. Math. Month.} {\bf 85} 2

\bibitem{rec1} Arnold B C, Balakrishnan N and Nagaraja H N, 1998 {\it Records} (New York: Wiley)

\bibitem{rec2} Nevzorov V B, 2001 {\it Records: Mathematical Theory} Translation of Mathematical Monographs {\bf 194} (Providence, RI: American Mathematical Society)

\bibitem{kr} Krapivsky P L and Redner S, 2002 {\it Phys. Rev. Lett.} {\bf 89} 258703

\bibitem{knuth} Graham R L, Knuth D E and Patashnik O, 1994 {\it Concrete Mathematics} (Reading, MA: Addison-Wesley)

\bibitem{feller} Feller W, 1968 {\it An Introduction to Probability Theory and its Applications} (New York: Wiley)

\bibitem{cox} Cox D R, 1962 {\it Renewal theory} (London: Methuen)

\bibitem{cox-miller} Cox D R and Miller H D, 1965 {\it The Theory of Stochastic Processes} (London: Chapman \& Hall)

\bibitem{gl} Godr\`eche C and Luck J M, 2001 {\it J. Stat. Phys.} {\bf 104} 489

\bibitem{dyson} Dyson F J, 1953 {\it Phys. Rev.} {\bf 92} 1331

\bibitem{alea} Luck J M, 1992 {\it Syst\`emes d\'esordonn\'es unidimensionnels} (in French) (Collection Al\'ea-Saclay)

\bibitem{ln} Luck J M and Nieuwenhuizen T M, 1988 {\it J. Stat. Phys.} {\bf 52} 1

\bibitem{golombdickman} Finch S R, 2003 {\it Mathematical Constants} (Cambridge: Cambridge University Press)

\nonum Weisstein W E, {\it Golomb-Dickman constant} Mathworld http://mathworld.wolfram.com

\bibitem{dickman} Dickman K, 1930 {\it Arkiv f\"or Mat., Astron. och Fys.} {\bf 22} 1

\bibitem{golomb} Golomb S W, 1963 {\it Bull. Amer. Math. Soc.} {\bf 70} 747

\bibitem{goncharov} Goncharov W, 1944 {\it Izv. Akad. Nauk SSSR} {\bf 8} 3

English translation: 1962 {\it Amer. Math. Soc. Transl.} {\bf 19} 1

\bibitem{shepp} Shepp L A and Lloyd S P, 1966 {\it Trans. Amer. Math. Soc.} {\bf 121} 340

\bibitem{gold} Goldie C M, 1989 {\it Math. Proc. Camb. Phil. Soc.} {\bf 106} 169

\bibitem{df} Derrida B and Flyvbjerg H, 1987 {\it J.Phys. A} {\bf 20} 5273

\bibitem{kesten} Kesten H, 1973 {\it Acta Math.} {\bf 131} 208

\bibitem{calan} de Calan C, Luck J M, Nieuwenhuizen T M and Petritis D, 1985 {\it J. Phys. A} {\bf 18} 501

\bibitem{vervaat} Vervaat W, 1979 {\it Adv. Appl. Prob.} {\bf 11} 750

\bibitem{maj} Majumdar S N and Ziff R M, 2008 {\it Phys. Rev. Lett.} {\bf 101} 050601

\bibitem{fb} Ferretti L and Bianconi G, 2008 arXiv:0804.1768 to appear in {\it Phys. Rev. E}

\end{thebibliography}
\end{document}